\def\year{2017}\relax
\documentclass[letterpaper]{article} 
\usepackage{aaai17}  
\usepackage{times}  
\usepackage{helvet}  
\usepackage{courier}  
\usepackage{url}  
\usepackage{graphicx}  
\usepackage{wrapfig}
\usepackage{algorithm}
\usepackage{algorithmic}
\usepackage{amssymb}
\usepackage{graphicx}
\usepackage{caption}
\usepackage{subcaption}
\usepackage{color}
\usepackage{multirow}

\begin{document}

%
\title{What's up with Privacy?: User Preferences and Privacy Concerns in \\ Intelligent Personal Assistants}
\author{Lydia Manikonda, Aditya Deotale, Subbarao Kambhampati \\ Arizona State University\\ \tt \{lmanikonda, adeotale, rao\}@asu.edu }

\nocopyright
\maketitle
\begin{abstract}
The recent breakthroughs in Artificial Intelligence (AI) have allowed individuals to rely on automated systems for a variety of reasons. Some of these systems are the currently popular voice-enabled systems like \emph{Echo} by Amazon and \emph{Home} by Google that are also called as Intelligent Personal Assistants (IPAs). Though there are raising concerns about privacy and ethical implications, users of these IPAs seem to continue using these systems. We aim to investigate why users are concerned about privacy and how they are handling these concerns while using the IPAs.  By utilizing the reviews posted online along with the responses to a survey, this paper provides a set of insights about the detected markers related to user interests and privacy challenges. The insights suggest that users of these systems irrespective of their concerns about privacy, are generally positive in terms of utilizing IPAs in their everyday lives. However, there is a significant percentage of users who are concerned about privacy and took further actions to address the related concerns. Some percentage of users expressed that they do not have any privacy concerns but when they learned about the ``always listening'' feature of these devices, their  concern about privacy increased. 
\end{abstract}

\section{Introduction \& Motivation} 
We define Intelligent personal assistant (IPA) as a system that is capable of learning the interests and behavior of the user and respond accordingly. Some of these systems that are always on and listening in the background are becoming ubiquitous. Despite the many possible privacy concerns raised by these IPAs, their popularity is continuing to soar. In the current market, there exist multiple IPAs -- \emph{Amazon Echo} (http://www.amazon.com/oc/echo/), \emph{Google Home} (https://madeby.google.com/home/), \emph{Microsoft Cortana} (https://www.microsoft.com/en-us/windows/cortana), \emph{Apple Siri} (www.apple.com/ios/siri/). However, due to the independent nature and existence in the physical form we primarily consider Amazon \emph{Echo} and Google \emph{Home} in our investigation. According to a recent analysis by Adobe Digital Insights\footnote{https://goo.gl/nCzjNt}, there exist 35.6 M users of voice-enabled speakers. Where, 70.6\% of them use Amazon Echo and 23.8\% of them use Google Home while the rest use other types of systems. Given the popularity of these IPAs and the raising concerns about privacy, now is the time to investigate how individuals are finding a trade-off between privacy concerns and the convenience and efficiency of the IPAs. 

This paper leverages the machine learning techniques along with the help of human coders to investigate three main research challenges 1) unique markers of IPAs that motivate individuals to use them; 2) privacy concerns associated with the usage of IPAs; 3)actions taken to address the privacy concerns. In order to address these challenges, we investigate data about IPAs from two main sources of information 1) reviews posted on the web and 2) responses to a survey. The digital version of word-of-mouth -- online reviews are powerful sources of information especially the prime factors of new product diffusion~\cite{lin2011effects}.  Alongside, surveys provide an in-depth understanding of the perceptions and experiences of individuals~\cite{rosentraub1981use}. This approach of utilizing both online and offline sources of information ensures that the patterns extracted from the data provides meaningful and trustworthy insights. Especially this provides a deeper understanding about the privacy issues of IPAs. 

\begin{table*}[h!]
\centering
\begin{tabular}{lccc}
\hline
& \textbf{Best Buy} & \textbf{Google Search} & \textbf{Amazon} \\
\hline
\textbf{Amazon Echo +ve} (26,402)  & 4880 & 912 & 20,610 \\
\textbf{Amazon Echo -ve} (4,833) & 256 & 67 & 4510 \\
\hline
\textbf{Google Home +ve} (5,495) & 4638 & 857 & 0 \\
\textbf{Google Home -ve} (655) & 514 & 141 & 0 \\
\hline
\end{tabular}
\caption{Distribution of the reviews across the two IPAs}
\label{tab:reviewstats}
\end{table*} 

\noindent\textbf{Summary of results:} Our analysis led to five key insights about the usage patterns of IPAs and how users balance privacy concerns. From the online reviews -- 1) Even though there is a significant percentage of negativity towards IPAs, users are highly positive towards using them in their daily lives; 2) There are larger percentage of concerns with regard to privacy for Amazon \emph{Echo} predominant from January 2017 compared to Google \emph{Home}. Interestingly, this is coincidental to the infamous court case of an Arkansas man accused of killing his friend in November 2015. In the trial of this case, the court ordered Amazon to hand over the recordings of \emph{Echo}. Responses to the survey suggested that -- 3) users grapple with seven different privacy issues. They are -- the device getting hacked (68.63\%), collecting personal information (16\%), listening 24/7 (10\%), recording private conversations (12\%), not respecting user's privacy (6\%), data storage repository (6\%), creepy nature of this device(4\%); 4) Even technically-savvy users are not completely aware of the fact that these devices are always listening. Once they they were informed of this fact, their concernes about privacy increased; 5) Users who mute their microphone expressed privacy as one of their main reason to mute the device. 

The results of our studies thus shed some light on the complex ways users are still wrestling with the tension between the utility of the IPAs and the privacy concerns they raise.  We hope these can help in some small way towards better design of IPAs. 

\section{Data Collection and Pre-processing}
The reports shared by Business Insider states that 35.6M users in the United States use voice-enabled speakers which include popular IPAs -- Amazon \emph{Echo} and Google \emph{Home}. Due to the popularity of these products, there are multiple sources of information about the features of these products, articles about comparing the products, reviews written online by the users of these products. We wish to utilize the wealth of information available online in the form of product reviews to investigate the privacy aspects of IPAs. The first task is to crawl the reviews and process them to create the final dataset used in this investigation. We crawl\footnote{Python crawler \url{https://docs.python.org/2/howto/urllib2.html}} the reviews from 1) Google Search, 2) Amazon (\url{amazon.com}) and 3) Best Buy (\url{bestbuy.com}). We obtain a total (includes both positive as well as negative reviews) of 31,235 reviews for Amazon Echo and 6,150 reviews for Google Home. For each review we crawled, there are multiple attributes attached with it -- rating, title, author, date posted, review text, \#comments, \#votes. The detailed statistics about the reviews are shown in Table~\ref{tab:reviewstats}.

After analyzing the reviews posted online, we designed a survey to address the open-ended questions emphasizing on the concerns about privacy while utilizing IPAs. To avoid introduction of any gender based biases in the responses, we developed the survey to reach both men and women. We obtained the IRB approval for this study. The participants in our survey must be 18 or older and are granted consent after reading a description of the study. Minors are not allowed to participate even with their parental consent. We recruited 51 participants who are students on a university campus and used IPAs previously or are currently using. The recruited students are earning their bachelors degrees in the STEM fields. As shown by the survey demographics in the Table~\ref{tab:surveystats}, 94.2\% of the recruited participants belong to the age group of 18-20 years old. The survey starts with questions about the type of product they use (either Amazon \emph{Echo} or Google \emph{Home}) and different factors associated with their interest in using this product. This set of questions are followed by questions about their concerns about privacy or security aspects along with their opinion on muting the microphone. At this section of the survey, the participants are aware of the default setting of always on and listening continuously for the trigger phrase. We then repeat the questions about their concerns about privacy and end the survey with an open-ended question about any feedback about changing the product to respect user's privacy. Two researchers read the responses to these open-ended questions posed in the survey and coded the data so as to answer the questions we posed earlier. This coding process was done separately and until both the researchers reached an agreement.


\begin{table}[h!]
\centering
\begin{tabular}{|l|p{0.55\linewidth}|}
\hline
\textbf{Age}  & 49 18-20, 2 21-24, 0 25-30, 0 30 or more \\
\hline
\textbf{Gender} & 45 Male, 6 Female\\
\hline
\textbf{Main usage} & Listening to Music(44), Personal Assistant(24), Controlling other devices(11), Getting Information like Weather(36) \\
\hline
\end{tabular}
\caption{Survey Demographics and Information}
\label{tab:surveystats}
\end{table}


Irrespective of working with the crawled online reviews dataset or the responses to the survey, the data should be pre-processed due to the considerable noise in terms of non-english characters, emoticons, informal language, etc. (c.f. \cite{baldwin2013noisy}). Without the processing of the data, the machine learning algorithms that are utilized in the latter part of the analysis are forced to work with a data that is unstructured and ambiguous. Due to these reasons the data is processed in such a way that the emoticons, non-english characters are removed and structure is imposed on this data. Also, since the reviews are crawled from multiple online sources, there is a possibility that we might've crawled certain reviews multiple times. 


\begin{figure*}[t!]
    \centering
    \begin{subfigure}[t]{0.45\textwidth}
        \centering
        \includegraphics[height=0.5\textwidth]{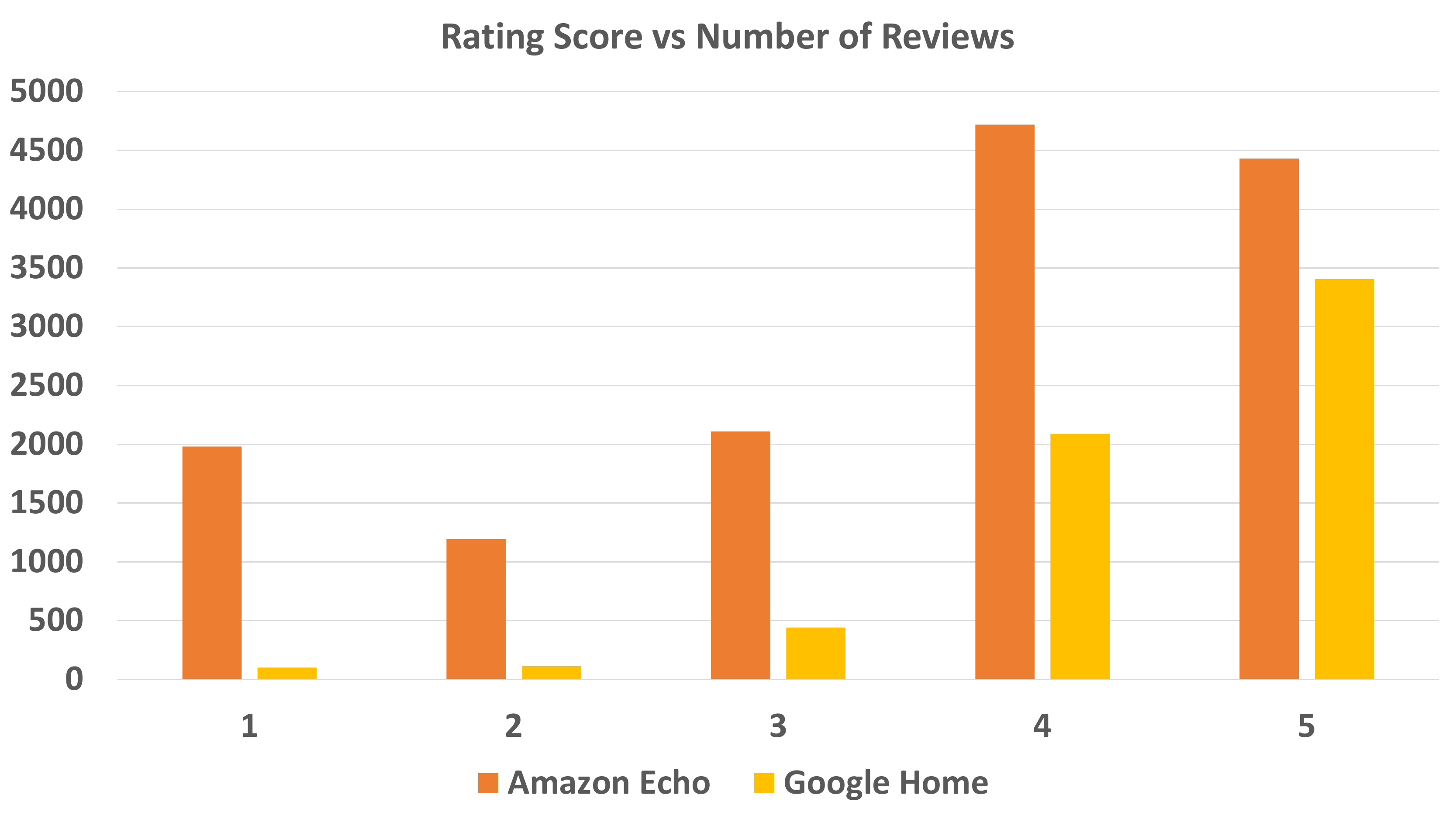}
        \caption{}
    \end{subfigure}%
    ~ 
    \begin{subfigure}[t]{0.45\textwidth}
        \centering
        \includegraphics[height=0.5\textwidth]{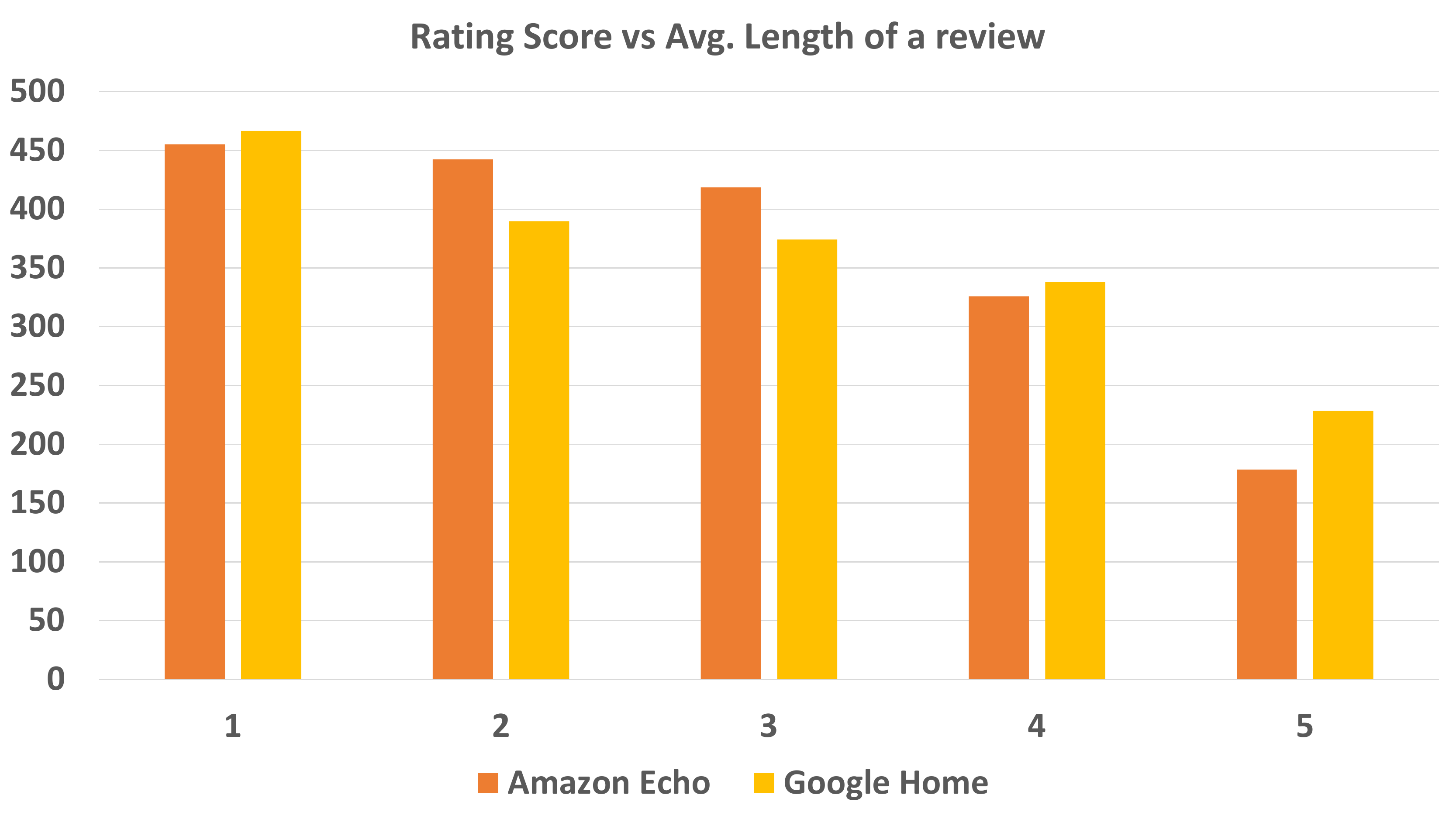}
        \caption{}
    \end{subfigure}
    \caption{ (a)Rating score vs the number of reviews for the corresponding score; (b) Rating score vs average length of the reviews for the corresponding score}
    \label{fig:ratings}
\end{figure*}

\section{Analyzing the Online Reviews}
We utilize the online reviews crawled for both Amazon \emph{Echo} and Google \emph{Home} in such a way that the positive reviews for both these devices are merged together in this section of the analysis unless specified. Similarly, the negative reviews are merged together for both the devices. We consider the 4-star rating and the 5-star rating as positive reviews and 1-star, 2-star and 3-star ratings as negative reviews. Table~\ref{fig:ratings} shows that the number of ratings for negative reviews are fewer than the positive reviews however, the average length of negative reviews is larger than the positive reviews.

\subsection{Reviews Seem to be Surprisingly Positive than being Negative}
Online expressions in the forms of reviews and ratings, are the artificial currencies for marketers to learn about the impact of their products on society. Detecting emotions from these reviews help shed light on the different aspects of IPAs. Specifically in the context of this paper, emotions could comprehend the interactions with IPAs especially their effect on users and their cognitive abilities. To detect emotions, we employ the psycholinguistic lexicon LIWC (http://liwc.wpengine.com/) on the text associated with the reviews we crawled. It considers the following 10 emotional attributes motivated from prior work on the public opinions about AI~\cite{ManikondaDK17}. The emotional attributes are: \emph{insight}, \emph{sad}, \emph{anger}, \emph{home}, \emph{negative emotion}, \emph{family}, \emph{cognitive mechanisms}, \emph{bio}, \emph{positive emotion} and \emph{sexual}. 

\begin{figure}[h!]
\small
  \begin{center}
    \includegraphics[height=1.5in]{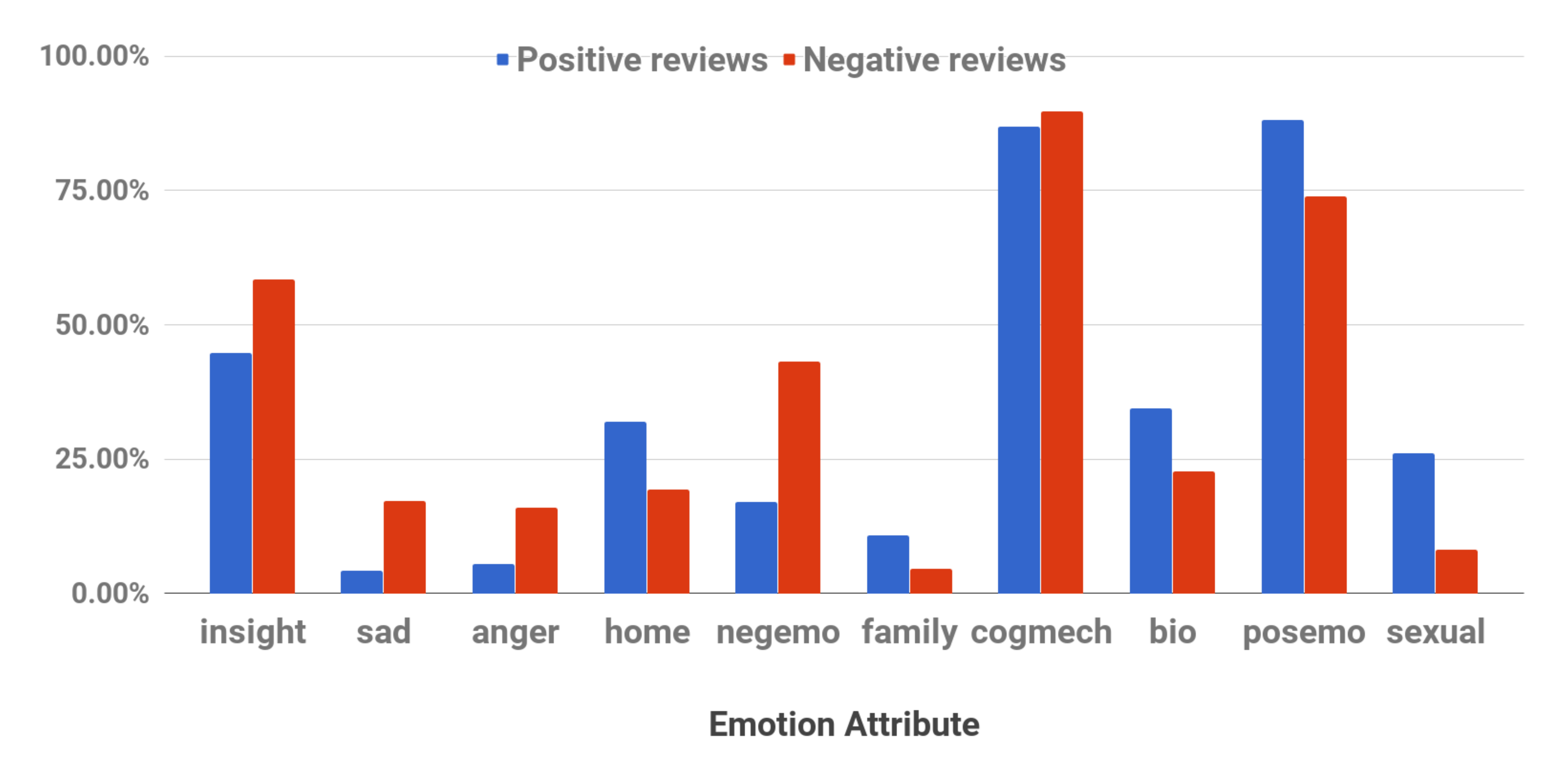}
  \end{center}
  \vspace{-5mm}
  \caption{10 prominent emotional attributes extracted using LIWC}
  \label{fig:liwcVals}
\end{figure}

Figure~\ref{fig:liwcVals} shows that since systems are intelligent in nature, it is interesting to observe that they are engaging the customers of these products in terms of their cognitive abilities (\textit{cogmech}=86.88\% (+ve) and 89.76\%(-ve)). Cognitive mechanisms can be described as a richer way of reasoning~\cite{tausczik2010psychological}. Negative reviews have larger percentage of insight words that may suggest the active reassessment of a theory. Using a large number of insight words especially in the negative reviews may suggest that customers focus on providing a detailed in-depth reviews justifying their ratings. The average length of the reviews shown in Figure~\ref{fig:ratings} support this observation that the length of negative reviews are longer than the positive reviews for the IPAs considered in this study. There is a significant percentage of words that belong to the categories of \textit{home}, \textit{family}, \textit{bio} and \textit{sexual} which focus on words related to different aspects of home, family, themselves and gender respectively. This suggests that users describe how their family or they evaluate this product. For example, one of the reviewer posts -- \emph{``My wife thought it was fun [...]'', ``My kids love the Pandora connection [...]''}. 

Emotions related to negativity -- \textit{sad}, \textit{anger}, \textit{negemo} are present in even the positive reviews apart from the negative reviews as shown in Figure~\ref{fig:liwcVals}. Irrespective of negativity towards these products, they do have a higher percentages of positivity (\textit{posemo}). For example, a negative review states -- ``\emph{Kind of neat, kind of cool but limited capability with Google play music. Bluetooth is nice but the range limitations make the Echo less than useful[...]}''. On the other hand, users who are giving high rating also expressed concerns about privacy. For example a reviewer who gave a 4-star rating said -- ``\emph{Works decent for voice queries, great speaker, the reason it isn't 5 stars is the privacy concerns, since it is always listening}''.

\subsection{Answers to the Questions are Limited but They are Great for Music} 
\begin{table*}
\centering
\begin{tabular}{|l |p{8cm}|}
\hline
\textbf{Review Type} & \textbf{Summarized Topics and their distributions} \\
\hline
\textbf{Positive Reviews} & Query search about news, weather, calendar, etc (31.6\%); Voice recognition and speaker capabilities (8.3\%); purchased or got it as a gift and how the family has fun playing with it (34.4\%); Easy integration with other devices or playlists (25.8 \%)\\
\hline \hline
\textbf{Negative Reviews} & Limited search features (63.8\%); wifi and network connection issues (14.2\%); contacted customer service for replacement or warranty and waste of money (22.0\%)\\
\hline
\end{tabular}
\caption{Topics extracted from the positive reviews and negative reviews}
\label{tab:topics}
\vspace{-5mm}
\end{table*}

In order to determine how and why users are utilizing IPAs in their everyday lives, discovering the latent topics focused by their reviews could be very helpful. Topics are primarily useful in terms of understanding the underlying semantic structure in the reviews. Even though emotion analysis provides a clear view on the attitudes of individuals towards IPAs, they do not necessarily specify the reasons for these emotions. These reasons can be captured by extracting the semantic latent topics. 

To conduct a latent topic detection, we utilize the popular Latent Dirichlet Allocation algorithm~\cite{Blei2003LDA} that mines the latent topics given a set of documents. We extract the top-4 latent topics from both positive and negative reviews separately. Once the topics are extracted, coding process is conducted by 2 human encoders to summarize the topics (summarized topics in Table~\ref{tab:topics}). These topics extracted describe the reasons why users either like or dislike the product. For example, topics extracted from negative reviews are about how these IPAs have limited understanding of questions posed by the users along with the technical faults of the devices. These definitely point to the fact that the speech recognition, relevancy of answers for the questions are definitely drawing negative criticism from users especially for Amazon \emph{Echo} not so much for Google \emph{Home}. On the positive note, the easy setup of IPAs, smart control of lights, ease of communicating with speech are key factors that are attractive to the users.  

\begin{figure*}[t!]
    \centering
    \begin{subfigure}[t]{0.5\textwidth}
        \centering
        \includegraphics[height=0.6\textwidth]{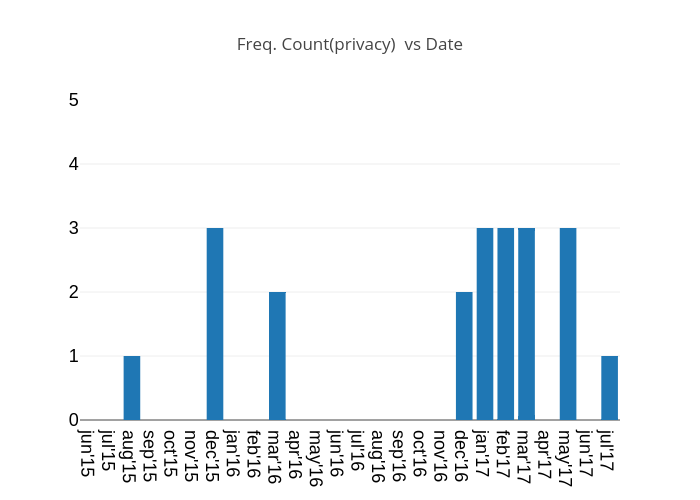}
        \caption{Privacy trends for Echo -ve Reviews}
    \end{subfigure}%
    ~ 
    \begin{subfigure}[t]{0.5\textwidth}
        \centering
        \includegraphics[height=0.6\textwidth]{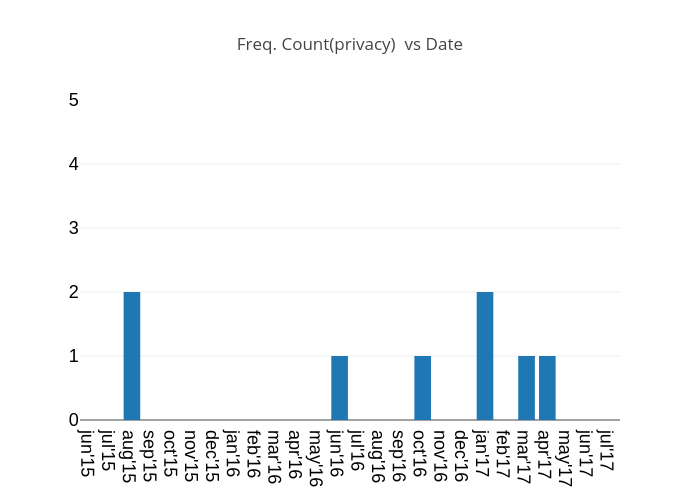}
        \caption{Privacy trends for Echo +ve Reviews}
    \end{subfigure}
     \caption{Privacy Trends over Time for Amazon \emph{Echo}}
     \label{fig:privacytrends}
\end{figure*}

\subsection{The Rise of Concerns about Privacy}
A traditional $n$-gram analysis cannot provide us with an accurate information about the semantic distances between words but are totally relied on the occurrence frequency. Co-occurrence or semantic proximity on the other hand is defined as the concept when two terms from a corpus co-occur in a certain order. In computer vision, co-occurrence matrix is utilized to measure the texture of an image that includes color and intensity. Leveraging this idea, co-occurrence matrix in a linguistic context is measured between pairs of words to extract the distinct markers of the IPAs -- Amazon \emph{Echo} and Google \emph{Home}. These words are represented in the form of vectors by the Word2Vec model. Given the advantages of Word2Vec models~\cite{Mikolov2013}, we envision that measuring the co-occurrence matrix in terms of the Word2Vec space will identify distinguishable markers between the IPAs that emotion analysis and topic analysis might not have discovered. Even though the performance of these approaches increase as the amount of training data (which is the set of reviews for a given category) increases, they detect meaningful co-occurrence patterns that are both semantically and syntactically related. 

Table~\ref{tab:cooccur} lists the top co-occurring words with the phrase "privacy issues" that are syntactically and semantically related. Based on these top co-occurring words shown in this table, Amazon \emph{Echo} received more concerns about privacy than Google \emph{Home}. One reason could be that \emph{Echo} has been existing since 2015 where as \emph{Home} was launched in 2016. The other reason could be the negative attention received by \emph{Echo} during the infamous court case in December 2016 (https://goo.gl/nmrF8J). We also speculate that since Google already has an extended presence in terms of providing email services, mobile operating system, etc., users might be comfortable even if \emph{Home} is always on. As described in the earlier section, positive reviews contain negative comments which we also see in this particular context. The words that are highly associated with ``privacy issues'' extracted from the positive reviews are very negative. Especially, the words \emph{disappointed}, \emph{lack}, \emph{trouble}, \emph{hate} represent a deeper level of frustration towards this product. When it comes to the words extracted from the negative reviews, some of the users are either unplugging the device or returning the product. On the other hand, Google \emph{Home} users associated the word ``returned'' with ``privacy issues''. 

\begin{table}[ht]
\small
\begin{tabular}{c|c|c|c}
\hline
\multicolumn{2}{c}{\textbf{\em Amazon Echo}} & \multicolumn{2}{c}{\textbf{\em Google Home}}\\ \hline
\emph{+ve Reviews} & \emph{-ve Reviews} & \emph{+ve Reviews} & \emph{-ve Reviews} \\ \hline
lack & failure & noise&returned \\
serious & security &problem & \\
conversation & serious & & \\
disappointed &unplugged  & & \\
problem & stuck & & \\
unplug & lack & & \\
issue & returning & & \\
trouble & sadly & & \\
hate &  & & \\
\hline
\end{tabular}
\caption{Top co-occurring words with the phrase ``privacy issues''; Co-occurrence patterns are extracted using Word2vec model trained on the reviews posted online.}
\label{tab:cooccur}
\end{table} 

\subsubsection{Temporal Trends of Mentions about Privacy:}
In this age of Twitter acting as a news media platform, individuals are gaining better opportunities to get exposed to personalized news about real world scenarios. Especially, due to the rising fears of machines taking over the world, scenarios where intelligent systems are playing a key role in the society are gaining more attention from the media as well as public. By utilizing the review ratings, we conduct a temporal analysis to identify interesting temporal trends (shown in  Figure~\ref{fig:privacytrends}). We noticed that since December 2016, there have been concerns about the aspects of privacy predominantly among the negative reviews. This pattern is more evident for Amazon \emph{Echo}. Interestingly, it is coincidental with the infamous court case in December 2016\footnote{https://goo.gl/nmrF8J} where the court ordered Amazon to hand over the information recorded by Echo. 

Some of the reviews after December 2016 are: ``I returned my Echo because of privacy concerns. Now judges are issuing subpoenas in court cases for Echo records [...]'' written in May 2017, ``[...]with the recent court revelations [...] A legal nightmare potential for anyone who values at least a modicum of civil rights or cares about not only their privacy, but any of their visitors as well'' written in January 2017, ``[...] very cool to most, but for me the scope of privacy intrusion is beyond anything I've even imagined. Alexa, I think I can manage well without you.'' written in March 2017. This may suggest the key role played by the news stories and movies on affecting the user concerns about privacy.  

\section{Implications from the Responses to the Survey}
51 participants were participated in the survey where they were asked a series of open-ended questions to understand the usability patterns and  privacy concerns associated with the usage of these products. The demographics of the survey are shown in Table 2. This set of participants comprises of 45 males and 6 females who are mostly from the age group of 18-20 years old. Except 1 participant, rest of the other are currently using Amazon \emph{Echo} at home. We summarize the key takeaways about privacy concerns from the responses to the survey.  

\subsection{Seven Types of Privacy Issues Concern Users}
68.63\% of users responded to the survey that they have concerns with regard to the privacy while utilizing Amazon \emph{Echo} in their daily lives. Each participant can list one or more concerns about the privacy. Two researchers independently aggregated these responses and coded them into seven categories of privacy concerns. 

\subsubsection{1. Hacking the device:}
16\% survey respondents mentioned that hacking or tampering their device through remote web attacks are very concerning. For some, hackers interfering with their device was the first concern they thought of: ``I hope hackers don't have access to my info[...]'. Others are concerned that they might be vulnerable to malware or cybersecurity breaches: ``[...] information can be intercepted by 3rd parties such as malware, cybersecurity breaches, etc''. 

\subsubsection{2. Collection of Personal Information:}
16\% of our survey participants mentioned that collection of their personal information by the device is a primary privacy concern. For most of these participants the way that the device collects their personal information including details about their credit card information are concerning. Some of the example responses are: ``\emph{collection of personal information}'', ``\emph{[...]I am more concerned about saying my credit card information}''.  

\subsubsection{3. Recording Private Conversations:}
12\% of our survey participants indicated that they are concerned about the device recording their private conversations. Some of the responses are: ``Maybe discussing something private and someone is listening'', ``I do not want it to record my conversations'', ``unknown individuals could be listening to my life and that is totally appalling''.

\subsubsection{4. Listening 24/7: }
10\% of survey respondents expressed their concern that this device is listening to them all the time. Some of the responses are: ``It listens to my life 24/7'', ``Knowing every little thing about me and someone constantly watching''. Few respondents mentioned that they are not very popular so they don't mind the device listening to them but there are times at which they don't want the device to keep on listening: ``I am not someone who is super well known but some times I just don't want the device listening to me''.

\subsubsection{5. Respecting the user's privacy:}
6\% of survey participants mentioned that the device should respect their privacy. They didn't specify much in details but they said their privacy should be of utmost consideration. Some of their responses are: ``[...] I know how important security is and I value being able to keep my information private''.

\subsubsection{6. Data Storage Repository:}
6\% survey respondents were curious about where and how their information is being stored and utilized by the company that manufactured this device. Some of their responses are: ``where they keep the information [...]'', ``need to know where my history and private information is stored''. 

\subsubsection{7. Creepy nature: }
4\% of our survey participants issued concerns about the creepiness of this device. Some of their responses are: ``Its a bit creepy [...]'', ``I do not like the fact that it can be used as evidence''.

\subsection{When users learned about the devices always listening, their privacy concerns increased}

The survey asks the participants to rate their level of concern on a 1-to-5 likert scale before and after they were told about the ``always on'' feature to measure their change of concerns about privacy. First, the users were asked (\emph{before\_score}) to rate their level of concern. This question was followed by another question about their awareness of the fact that the device is always on and listening. 19.6\% of participants mentioned that they are not aware of this fact and then they were informed about this listening aspect of the device. The users were then asked (\emph{after\_score}) to rate their level of concern after knowing this additional information. We analyzed how the rate of concern for the 19.6\% of participants who are not aware of this fact about the device. To our surprise the rate of concern went up high by 50\% of the respondents who mentioned that they do not know the device is always on and listening. 



We dug a little deep into the responses of users whose scores were increased (for example, 1 to 3 and 2 to 5) especially the other factors of awareness about the product. These participants also mentioned that they  are not aware of the available option that they can mute the microphone. Some of these users also suggested that factors like this should be specified to the users upfront. This shows that the technically-savvy users are not completely aware of the fact that these devices are always listening but got alarmed after learning about this fact.

\subsection{Privacy is listed as one of the reasons to mute the microphone}
Many survey respondents expressed their concerns that the device is constantly listening to their conversations 24/7. 47\% of the respondents mentioned that they mute their microphone on this device and 29.4\% of them do not mute the microphone. The remaining 23.6\% survey respondents do not know that the option of muting their microphone is available. Then we asked the respondents their motivation to mute their microphones. Their aggregated responses suggest these reasons below:   
\begin{itemize}
\item background conversations
\item when not in use
\item privacy concerns (for example, someone responded that ``watched Snowden movie and made me weary of this stuff'' )
\item responds to questions when not needed
\end{itemize}

\section{Related Work}
Intelligent systems are becoming part of our lives as more and more individuals are relying on them for driving~\cite{kim2013parallel}, work management~\cite{ferguson1998trips,myers2007intelligent}, time scheduling~\cite{zhang2013cobi,manikonda2014ai}, information and email organization~\cite{mitchell1994experience,dumais2016stuff}, and responding to factual questions~\cite{brill2002analysis,ravichandran2002learning}. These systems leverage AI to learn and reduce the mistakes made by users due to cognitive overload~\cite{myers2007intelligent}. In this context, there are multiple debates about whether AI that is primarily used by these intelligent systems is a blessing or curse to the society. One of our previous works~\cite{ManikondaDK17} investigated the public perceptions of individuals about AI. These individuals comprise both AI experts and common users on Twitter (a popular micro-blogging platform) who are sharing their personal opinions and information about AI. However, we believe that researching the interactions of individuals with the intelligent systems has greater potential to understand the impact of AI on the society to a certain level especially the privacy concerns that come along with that~\cite{czibula2009ipa}. 

The digital version of word-of-mouth -- online reviews are becoming the major source of information especially a prime factor of new product diffusion for potential buyers~\cite{lin2011effects}. Existing literature~\cite{park2007effect,lee2009online} focusing on the impact of reviews state that better reviews led to improved sales of products. Also, surveys provide an in-depth understanding of the perceptions and experiences of users~\cite{rosentraub1981use}. In spite of the fact that the aspects of privacy with respect to intelligent systems are becoming popular, it has attracted relatively less attention from the research community. In this paper, using the online web reviews and a survey, we present a qualitative and quantitative analysis to study the privacy aspects of IPAs.  

\section{Conclusions}
We used online reviews along with the survey responses from the users of IPAs to draw the following conclusions: 1) People are overall very positive about these devices and comfortable to use them;  2) However, there are multiple privacy issues; 3) Many users expressed that they are aware of the privacy concerns and they do take actions: a) returning the product; b) muting the microphone; c) limiting  the usage -- example only using to set alarms; 4) Even technically-savvy  users (from our survey) are not completely aware of the fact that these devices are always on and listening. Once they learned that these devices are always on and listening, their rate of concern increased; 5) Users who mute their microphone expressed privacy as one of their main reasons to mute the device; 6) The temporal trends suggest that may be the media, in the form of news and movies are making users aware of the privacy concerns associated with these devices. Most of the respondents suggested that the privacy issues can be addressed by making these devices transparent. We hope that the results of our investigation highlights some of the complex ways users are struggling to find a balance between using these devices and the associated privacy concerns. We hope that these insights could help better design the IPAs with more transparency. 

\bibliographystyle{aaai}
\bibliography{references}

\end{document}